\documentclass[aps,prl,twocolumn,groupedaddress,amssymb]{revtex4}
\usepackage{graphicx,latexsym}
\begin{document}
\def\be{\begin{equation}}
\def\ee{\end{equation}}
\def\ba{\begin{eqnarray}}
\def\ea{\end{eqnarray}}
\title{Metallic-to-insulating transition in disordered graphene monolayers}

\author{M. Hilke}
\affiliation{Dpt. of Physics, McGill University, Montr\'eal, Canada H3A 2T8}

\begin{abstract}
We show that when graphene monolayers are disordered, the conductance exhibits a metallic-to-insulating transition, which opens the door to new electronic devices. The transition can be observed by driving the density or Fermi energy through the mobility edge. At the Dirac point the system is localized, whereas at higher densities there is a region of metallic behavior before the system becomes insulating again at higher densities. The region of metallic behavior depends on the disorder strength and eventually vanishes at high disorder. This result is quite unexpected since in square lattices, scaling theory predicts that this metallic region does not exist in two dimensions, in contrast to graphene, where the lattice is a honeycomb.
\end{abstract}

\maketitle
In most active electronic devices, the conductivity can be tuned from conducting to insulating by using a gate. With the recent discovery of graphene monolayers \cite{novo04}, and their potential for electronic devices \cite{geim07}, it is important for the conductivity to change substantially with the gate voltage. However, while in clean graphene monolayers, the conductivity changes almost linearly with gate voltage, the {\em off} or minimum conductivity is still relatively large (of the order of the unit of conductance $e^2/h$)  \cite{novo05,Mac06,Castro06}.

\begin{figure}[ptb]
\vspace*{-1cm}
\begin{tabular}{c}
\includegraphics[width=2in]{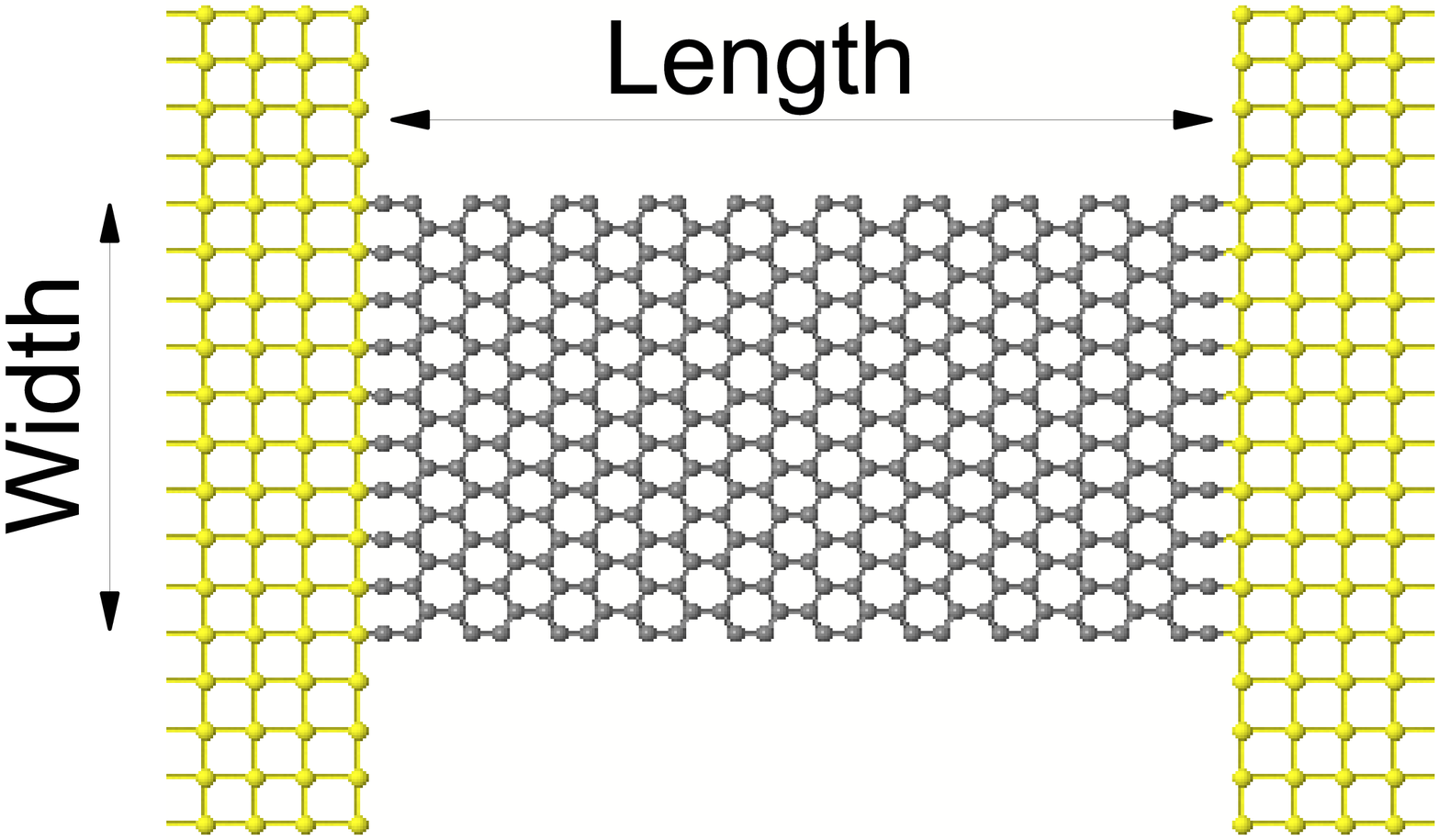}\\ \includegraphics[width=2.8in]{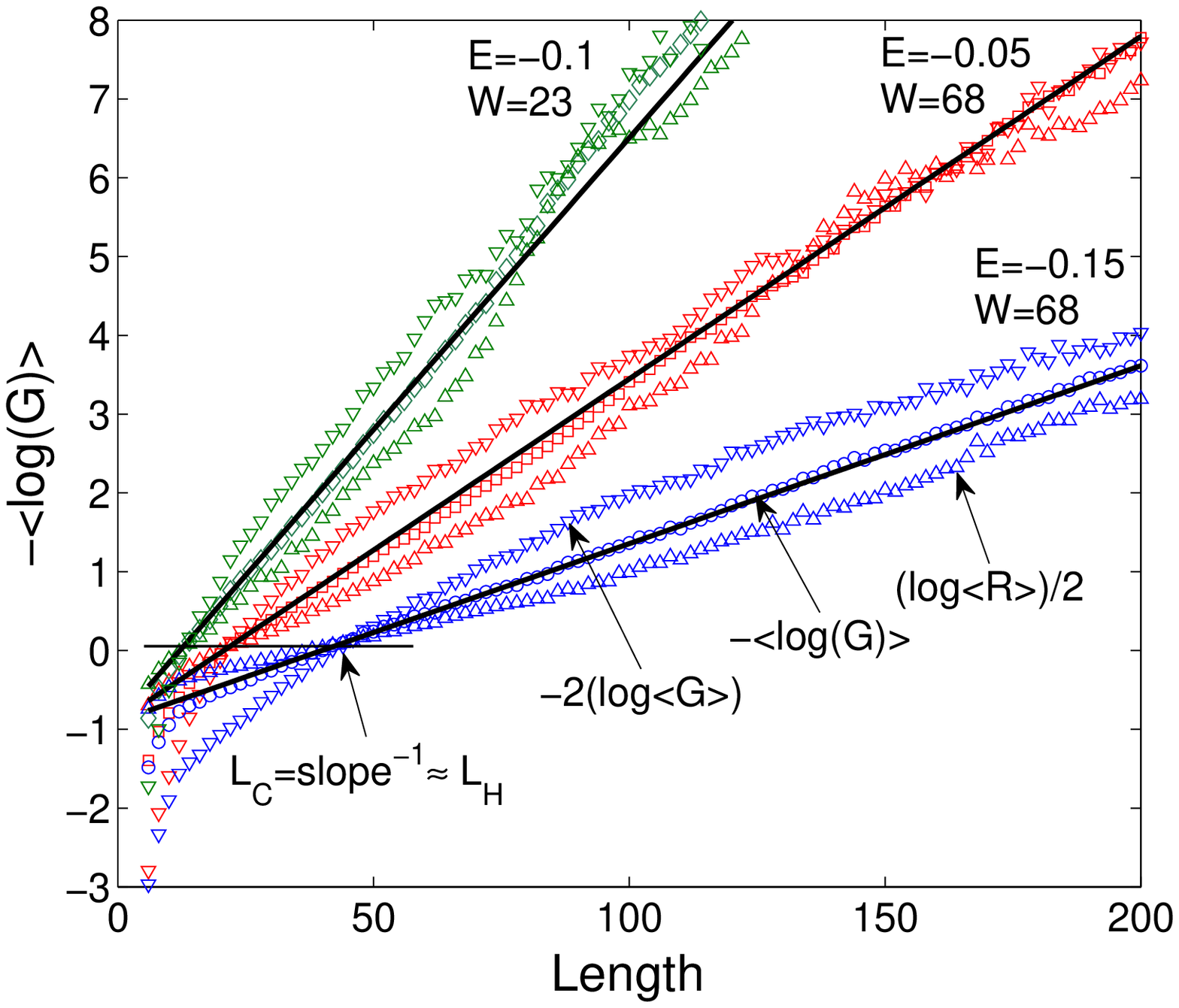}%
\end{tabular}
\caption{Top: Image of a graphene nanoribbon with two attached contacts. Main graph: Length dependence of $-\langle \log (G)\rangle$, $-2\log\langle G\rangle$ and $\log\langle R\rangle/2$ for different energies and widths. The average is taken over 200 configurations and $V=1.5$. The solid lines represent linear fits of the data for $L\gg L_C$, the slope of which gives the inverse localization length $L_C$. The horizontal line illustrates the value of $L_H$. $L$ and $W$ are given in units of graphene atoms.}%
\label{lnT_Ldep}%
\end{figure}

Here we consider the situation of graphene nanoribbons (with a honeycomb structure) of width $W$ and length $L$ contacted by two large normal metals, assumed to have a square lattice. This leads to a typical two-terminal configuration as shown in figure \ref{lnT_Ldep}, where the contacts are assumed to be perfect. The disorder potential is taken to be only onsite and uncorrelated. This is identical to the large body of work on Anderson localization in tight binding models \cite{Anderson58}, which show a metal-insulator transition in dimensions strictly higher than 2 \cite{gang4,3Da,3Db, 3Dc}. In two dimensional systems with a square lattice no metal-insulator transition exists for non-interacting electrons \cite{gang4,Schreiber92}, except if correlations in the disorder are present \cite{Hilke03}.

In graphene, with its honeycomb lattice, the band structure was first studied theoretically by Wallace \cite{Wallace47} using a tight binding Hamiltonian, where $t\simeq 3$eV is the hopping energy between two nearest neighbors and $V_n$ the onsite energy. At the band edges, the band structure is very similar to the band edge of square lattices. At the center of band, however, and for $V_n=0$ Wallace showed that there is a linear dispersion at the band center (Dirac point) at two points in the reciprocal space leading to a two-valley degeneracy. This has important implications on scattering properties, such as a suppression of intra-valley backscattering. The linear dispersion close to the Dirac point leads to dramatic new physics, such as an anomalous quantum hall effect and Dirac fermions \cite{CastroRev}. It is the absence of a gap at the Dirac point, which causes the conductivity to not vanish at that point \cite{novo05,Mac06,Castro06}, which is a potential roadblock for applications in active electronics.

Disorder plays an important role in graphene devices and can be due to ripples \cite{ishig07,chen08-2}, defects in the substrate \cite{hwang07} and surface effects, such as partial hydrogenization \cite{graphane}. To reduce disorder scattering, suspended graphene devices \cite{Stormer08} have been considered as well as the use of other substrates \cite{Yu09}.

{\it Localization in graphene nanoribbons}: Early numerical studies in disordered honeycomb lattices, were limited to the Dirac point and showed that the states are localized at this point \cite{Schreiber92}. These results were obtained by evaluating the localization length ($L_C$) from the smallest Lyapounov exponent of a finite width ($W$) ribbon using a transfer matrix approach. $L_C$ is then studied as a function of $W$. Using scaling arguments \cite{3Dc}, localized states are identified when $L_C/W$ decreases with $W$, whereas extended states are characterized by $L_C/W$ increasing with $W$. The point where there is no dependence is then inferred to as the critical point, where the localization-delocalization transition occurs. In two dimensional square lattices, only localized states were found for all energies \cite{2D}.

Here we use a very similar approach, but instead of considering a transfer matrix, we evaluate the two-terminal conductance of the system, which allows us to directly evaluate the transport properties of the system. For a given width of the graphene nanoribbon, we evaluate the zero-temperature two-terminal conductance ($G$) of the disordered graphene attached to metallic (square lattice and non-disordered) leads using an iterative Green's function technique. Since the conductance depends on the given disorder configuration we consider a configurational average $\langle\cdot\rangle$ by averaging over many disorder configurations. We used two disorder distributions, either uniform with $-V/2<V_n<V/2$, or binary ($V_n=\pm V/\sqrt{12}$) both characterized by $\langle V_nV_m \rangle=\delta_{n,m}V^2/12$. The two distributions give identical results, in contrast to one dimension, where large distribution dependent ensemble averaged conductance fluctuations exist \cite{Hilke08}. We assume the leads to be disorder free and infinitely long and much wider than the disordered graphene rectangle.

In the presence of disorder, the conductance will eventually vanish exponentially with the length of the system. This is illustrated in figure \ref{lnT_Ldep}, where we show  $-\langle \log(G)\rangle$ for different values of the Fermi energy in units of $t$. Because of the symmetric band structure around the Dirac point, all results are symmetric around $E=0$. Two regimes can be identified: (i) the ballistic regime, when $L\ll L_C$ and (ii) the localized regime, where $-\langle\log(G)\rangle\sim L/L_C$  for $L\gg L_C$. In the ballistic regime ($L\ll L_C$), the conductance is dominated by mesoscopic conductance fluctuations \cite{UCF1,UCF2}, where $\delta G\ll G$, hence
\be -\langle\log(G)\rangle\simeq -\log\langle G\rangle\simeq\log\langle R\rangle.\ee 
$R=1/G$ is the two terminal resistance of the device and $\delta G$ is the standard deviation of $G$. In the localized regime ($L\gg L_C$), on the other hand, where the conductance vanishes exponentially with the length of the system, we have $\delta G\gg G$. Because of the statistical properties of the conductance of a quasi-one dimensional system, this yields \cite{Pendry94} \be -\langle\log(G)\rangle\simeq -2\log\langle G\rangle\simeq\log\langle R\rangle /2\simeq L/L_C-\alpha. \ee $\alpha$ is a parameter, which is close to unity for $G$ in units of $2e^2/h$ as shown in figure \ref{lnT_Ldep}. Relation (2) becomes exact in the limit where $L\rightarrow\infty$. Hence, the localization length can be extracted using any of the average transport quantities ($\langle\log(G)\rangle$, $\langle G\rangle$ or $\langle R\rangle$), but the convergence of $\langle\log(G)\rangle$ is much faster as shown in figure \ref{lnT_Ldep}.

While equation (2) requires $L\gg L_C$ in order to extract $L_C$ accurately, an approximate $L_C$ can also be obtained by looking at the crossover from ballistic to localized, which will happen when equations (1) and (2) both hold, i.e., when each term is zero. This corresponds to $\delta G\simeq G$, since $\delta G$ is close to one in units of $2e^2/h$ \cite{UCF1}. Hence, defining $L_H$ as the length which minimizes $\min\{\langle\log(G(L))\rangle^2+(\log\langle G(L)\rangle)^2+(\log\langle R(L)\rangle)^2\}\Rightarrow L_H$, we obtain a length, which characterizes the crossover from ballistic to localized. It turns out that for all values of interest, $L_H\simeq L_C$ within 10\%.

{\it Scaling behavior:} In order to determine if a state is exponentially localized in the two dimensional limit, we need to evaluate $L_C/W$ as a function of $W$. $L_C$ is always finite in the quasi one-dimensional limit, since all states are localized. For the two-dimensional case, $L_C/W$ is the relevant quantity, since ultimately we are interested in the average conductivity $\sigma=\langle G\rangle L/W$ for infinite width and length. By leaving the aspect ratio constant and equal to one, i.e., $L=W$ when taking the limit to large sizes, we obtain $\sigma\sim e^{-W/2L_C}$, since $\langle G\rangle \sim e^{-L/2L_C}$ from equation (2). Hence, if with increasing width $L_C/W\rightarrow 0$ this leads to $\sigma \rightarrow 0$ and we have an insulator. On the other hand, if $L_C/W\rightarrow \infty$, for increasing width, this implies that there are no exponentially localized states and we refer to this state as metallic.

\begin{figure}[ptb]
\begin{tabular}{cc}
\includegraphics[width=1.7in]{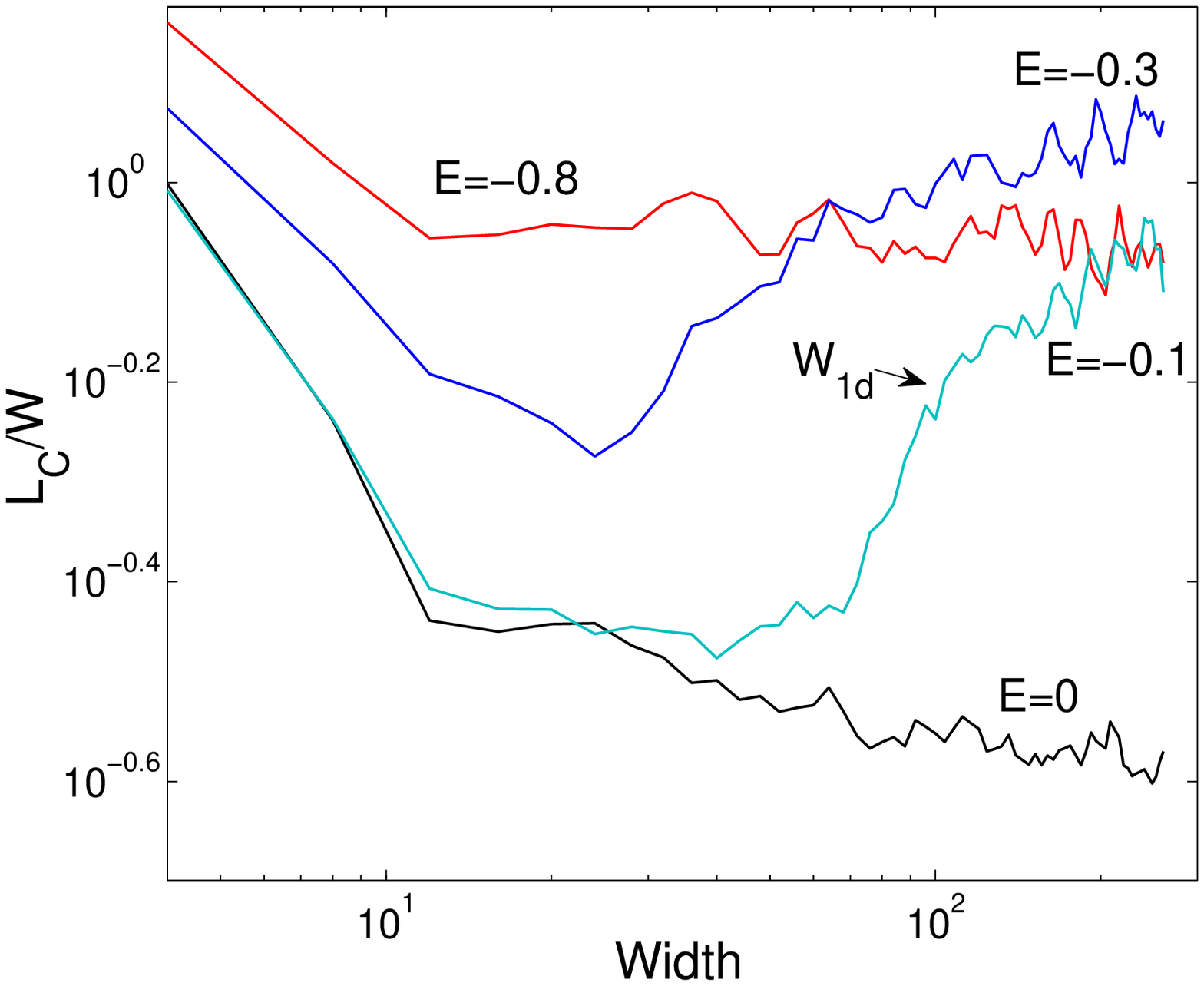} & \includegraphics[width=1.9in]{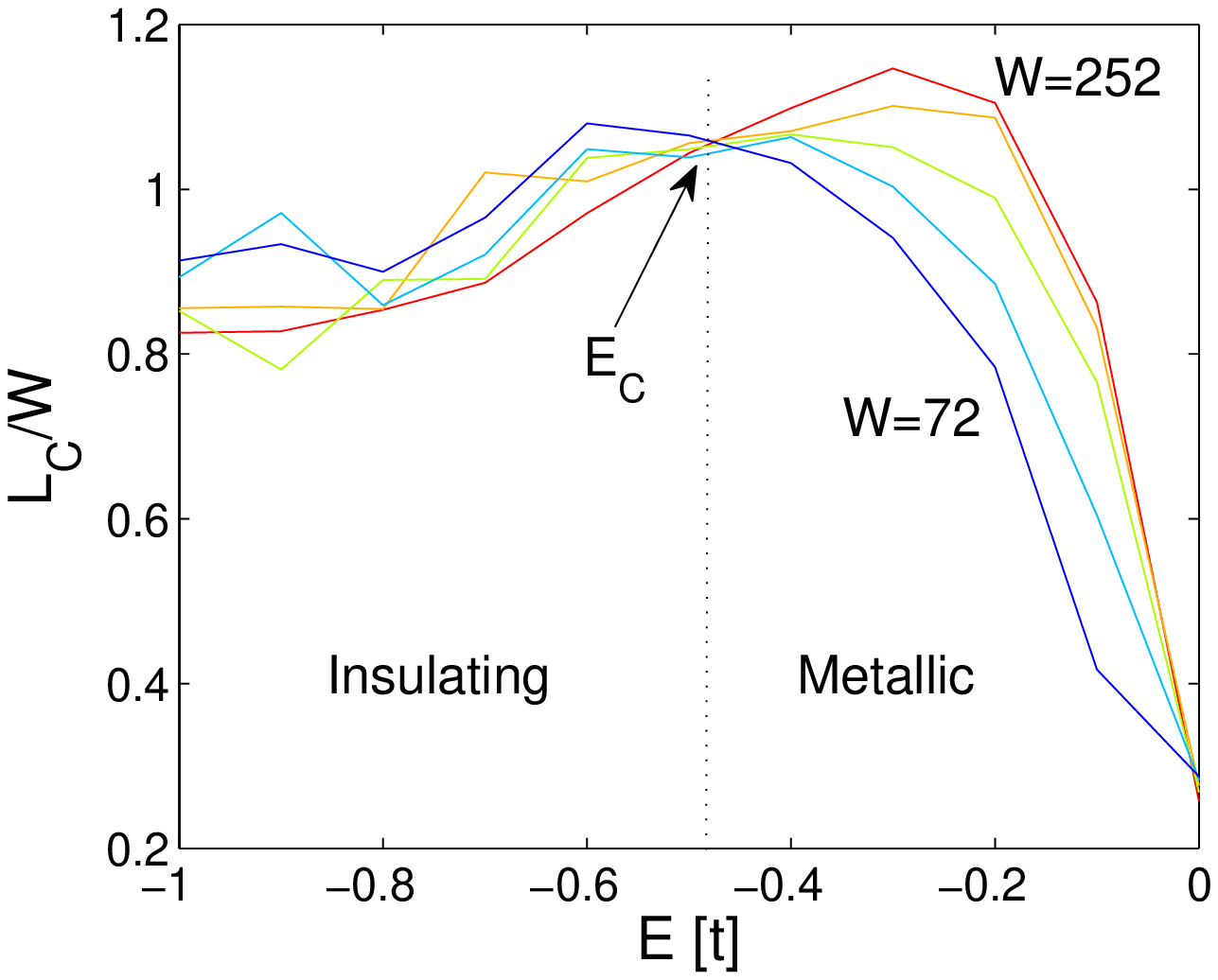}%
\end{tabular}
\caption{Left: The dependence on width of the ratio $L_C/W$ for different values of the energy. Right: The dependence of the ratio $L_C/W$ on energy for different values of the width. The energy $E_C$ labels the critical energy at which the width dependence of the ratio changes sign, indicative of a metallic-to-insulating transition.}%
\label{LcW2}%
\end{figure}

We now apply this analysis to our graphene device and show the results for $L_C/W$ in figure \ref{LcW2}. We observe that at the Dirac point, the ratio $L_C/W$ monotonously decreases with the width, which implies that the system is insulating at the Dirac point. This is consistent with earlier results \cite{Schreiber92,Mirlin06,Brouwer07}. More interestingly, away from the Dirac point, the dependence of $L_C/W$ is more complicated and the dependence becomes non-monotonous. Indeed, at small widths the ratio decreases at first, before increasing again at larger values of the width, which is a signature for metallic behavior. This increase of $L_C/W$ only occurs in a small window of energies, between $E=0$ and $E=E_C$ as shown in figure \ref{LcW2}. This is typical for metal-insulator transitions as seen in three dimensions \cite{3Dc}. However, this is quite unexpected in two dimensions without a magnetic field. Indeed, numerical studies on disordered systems with square lattices show no metallic behavior \cite{Schreiber92}, i.e., $L_C/W$ always decreases with size.

Interestingly, the continuous increase of $L_C/W$ over a wide range of widths close to the Dirac point, which is a signature for a metallic behavior, is correlated with the existence of additional channels. Indeed, for a non-disordered system close to the Dirac point, there is only one quantized transport channel, leading to a quantized conductance of $2e^2/h$ \cite{Castro09} if the width of the nanoribbon is smaller than $W_{1D}= 2\pi \hbar v_F/E_F$, where $v_F$ is the Fermi velocity and $E_F$ the Fermi energy. The value of $W_{1D}$ is shown in figure \ref{LcW2} for $E_F=-0.1t$ and is correlated to a jump in $L_c/W$. This is quite different from the square lattice case, where the addition of another channel is always correlated with a trough of $L_C$ \cite{Roemer04} and suggests the two situation to be very different in nature and indicates that the reason for this delocalization is intimately related to the linear dispersion, which leads to a suppression of intra-valley back-scattering. In terms of the beta function, Nomura and coworkers argued for the topological delocalization of two-dimensional massless Dirac fermions \cite{Nomura07}.

\begin{figure}[ptb]
\begin{tabular}{c}
  \includegraphics[width=2.5in]{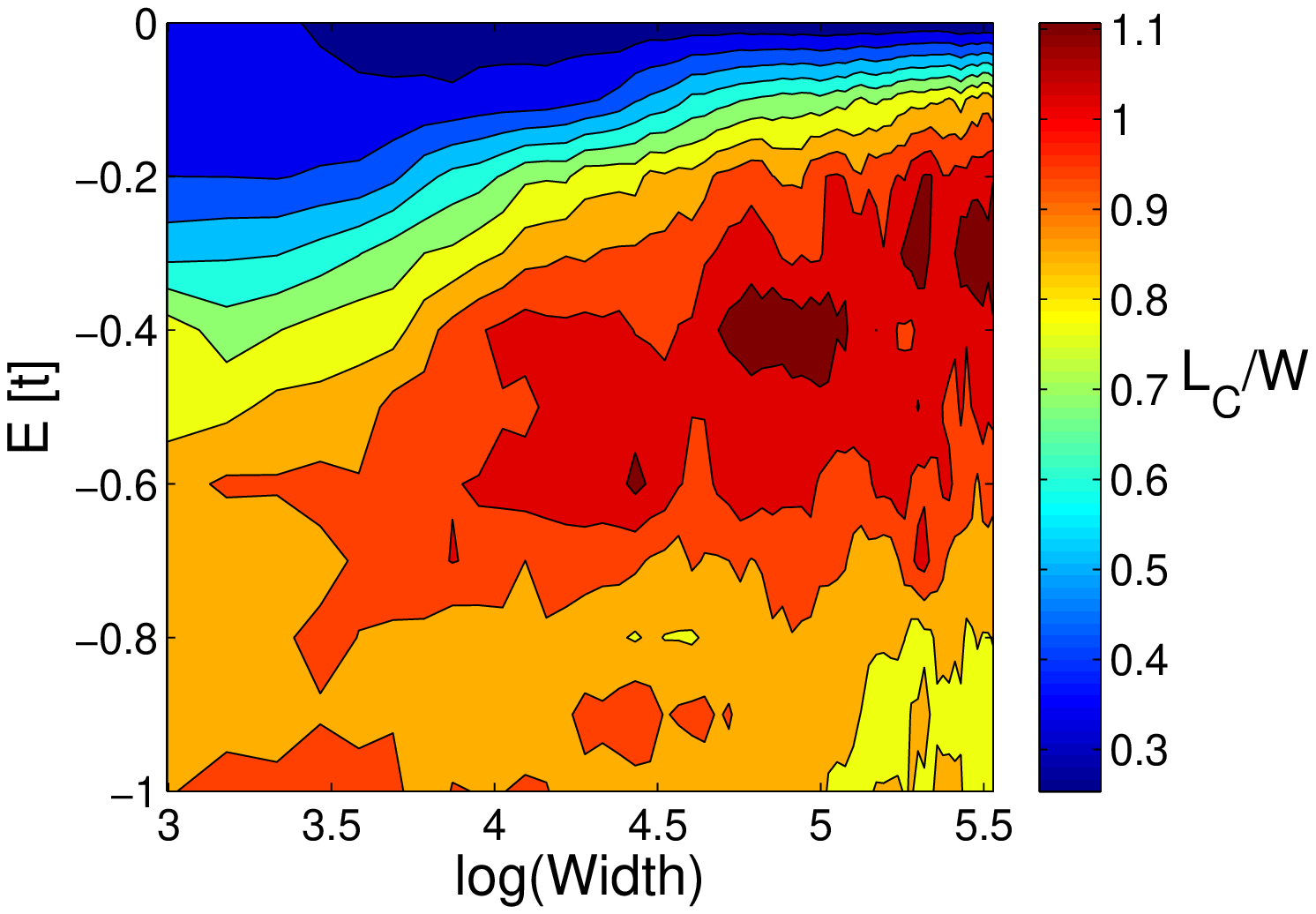}\\
  \vspace{-.2cm}\includegraphics[width=1.7in]{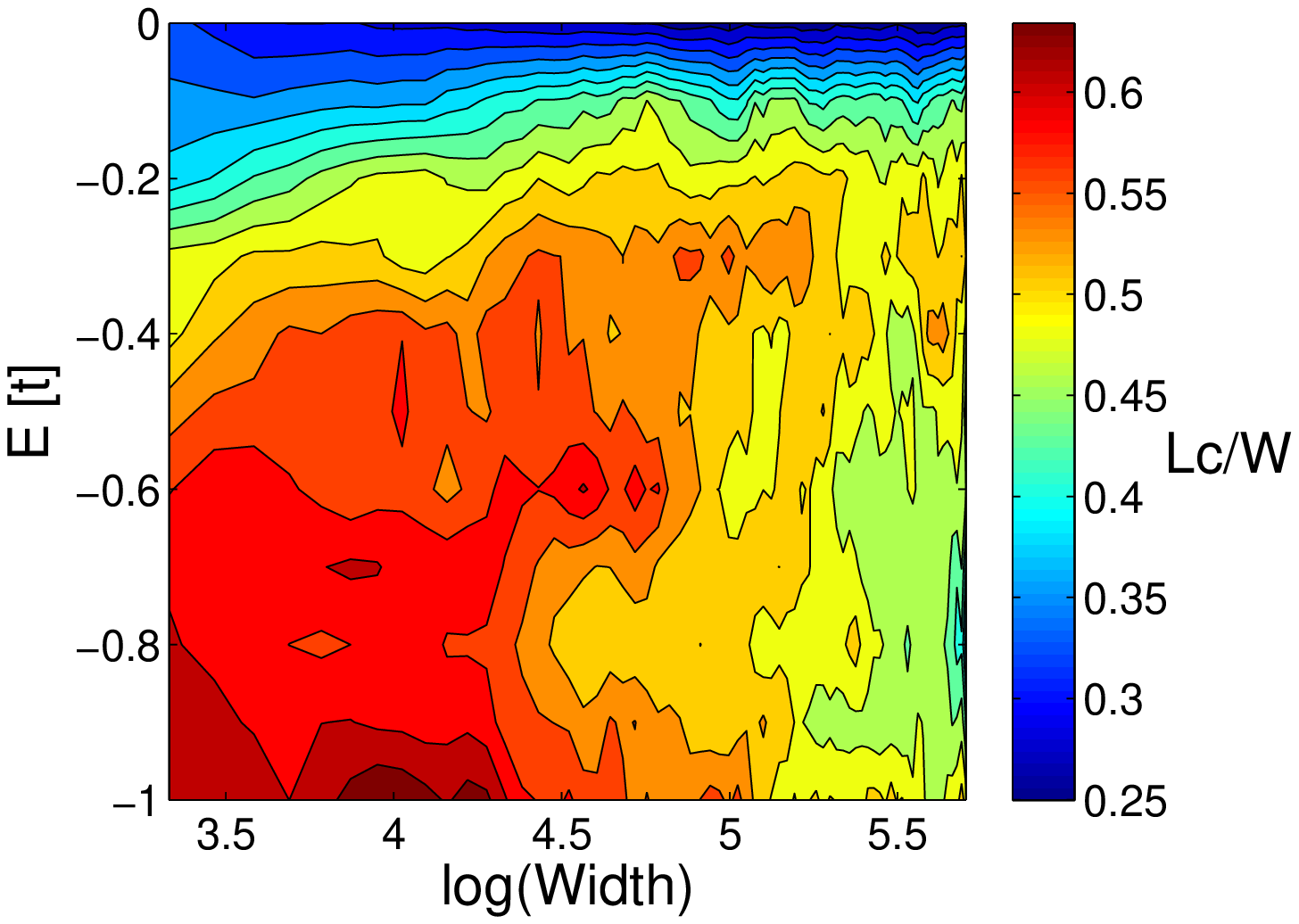}\hspace*{.2cm}
  \includegraphics[width=1.7in]{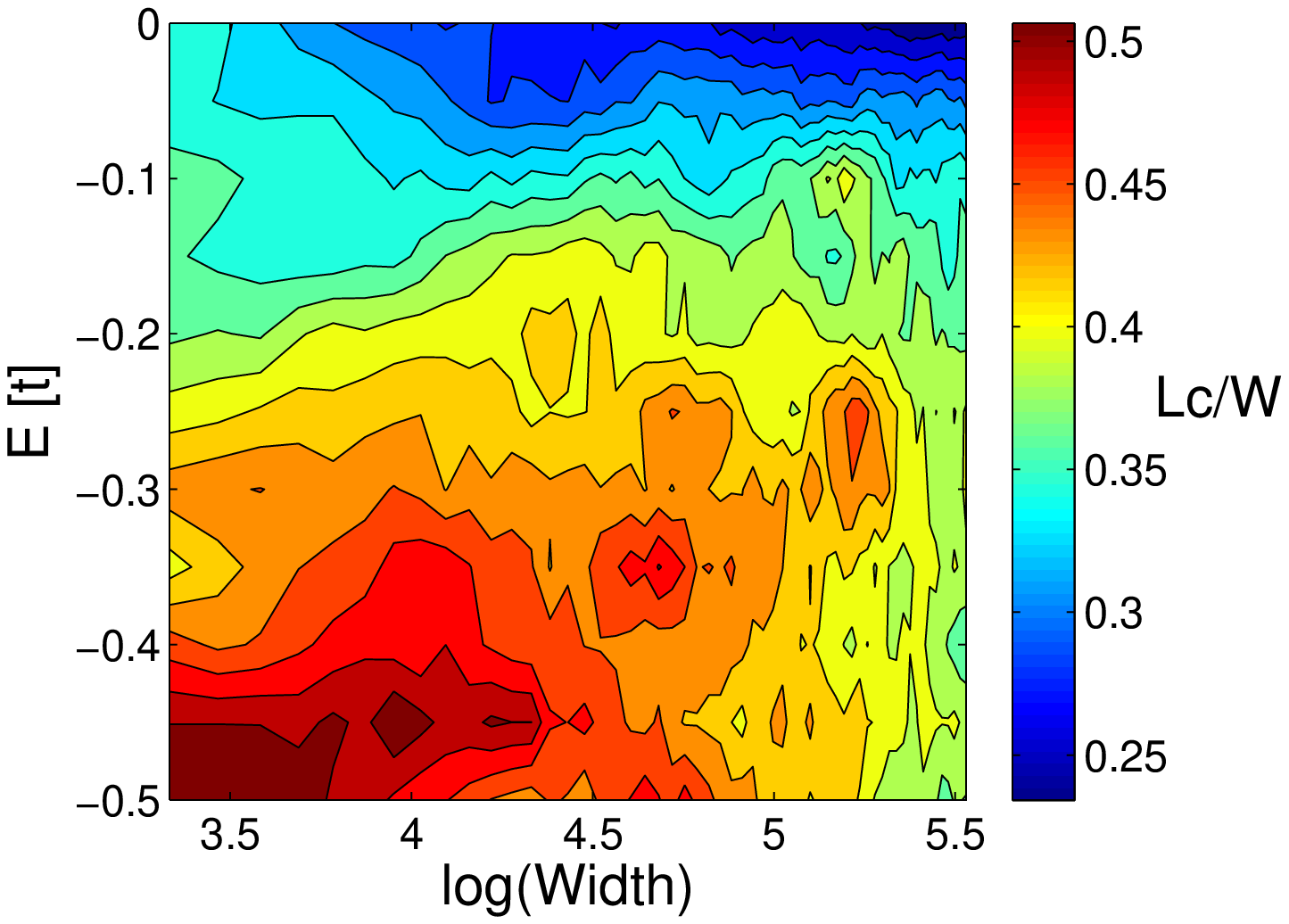}
  \end{tabular}
  \caption{Contour plots of the ratio $L_c/W$ as a function of the energy and width. Top graph is for $V=1.5$, bottom left for $V=2$ and bottom right for $V=2.2$.}
  \label{Contour}
\end{figure}

A more detailed analysis of the dependence of $L_C/W$ is provided by the contour plots shown in figure \ref{Contour}. For $V=1.5$ the phase space in energy and width where $L_C/W$ increases moves towards the Dirac point at $E=0$, yielding an increase in $L_C/W$ as a function of $W$ for an energy close to the Dirac point. In contrast, the contour plot for $V=2.2$ shows that for all energies $L_C/W$ decreases with the width of the system. This implies that all states are exponentially localized for $V=2.2$. The case of $V=2$ is interesting since it corresponds to the crossover between the two behaviors. Indeed, for $V=2$ and close to the Dirac point, the contour lines of constant $L_C/W$ are almost constant in energy, indicative of a critical behavior, in contrast to the $V=1.5$ case, where the contour lines decrease in energy with increasing $W$, which is opposite to the $V=2.2$ case.

\begin{figure}[ptb]
\includegraphics[width=3in]{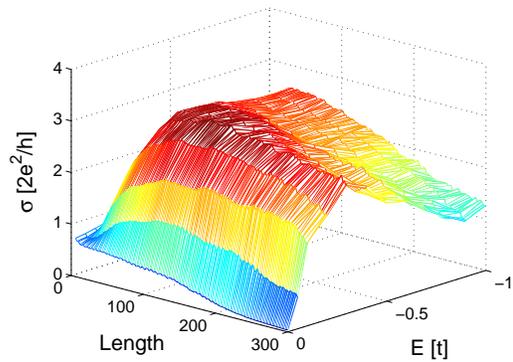}\\%
\caption{The dependence of the conductivity on energy and length of a nanoribbon 272 atoms wide for $V=1.5$.}%
\label{sigma}%
\end{figure}

The most interesting consequence of the existence of the metallic-to-insulating transition for disorder strengths smaller than $V\simeq 2$ is the possibility to have an {\it on/off} ratio as a function of density (gate voltage), which is infinite when taking disorder into account. In order to illustrate this point, we evaluated the average resistivity $\rho=\langle R\rangle \sqrt{3}W/4L $ for $V=1.5$ and plotted its inverse ($\sigma$) in figure \ref{sigma} as a function of energy and length of the nanoribbon for a fixed width of W=272 atoms. The numerical coefficient in determining $\rho$ is the geometrical factor associated with the way we defined the atoms for our numerical implementation. Under our scheme, the total number of graphene atoms is given by $W\times L$. We used $\langle R\rangle$ and not $\langle G\rangle$, because in most experiments $\rho$ is measured as a function of density using a fixed current.

For the maximum length (L=300), the conductivity vanishes at $E=0$ due to localization at the Dirac point. Moving away from the Dirac point the conductivity increases with energy due to the metallic behavior, before reaching a maximum, which leads to a diverging $\sigma_{\mbox{max}} / \sigma_{\mbox{min}}$ as a function of energy (density) for a given geometry and disorder strength. At even higher energies, the conductivity decreases again due to localization. This behavior is consistent with some recent experiments, where a similar behavior has been observed by photo-emission, consistent with a metal to insulator transition \cite{Bostwick09}. In epitaxial graphene, a metal to insulator transition was also observed by molecular doping \cite{Zhou08}. Several authors have computed the inverse participation ratio to show the existence of a gap close to the Dirac point \cite{Naumis07}. In a related work, the density of states was computed numerically and found to be consistent with a localization-delocalization transition close to the Dirac point \cite{Amini09,Fehske09}.

Summarizing, we have shown that there is a metallic-to-insulating transition in disordered graphene monolayers, which happens close to the Dirac point and leads to a window of energy, where a metallic behavior exists. This metallic behavior exists only for small enough disorder ($V\lesssim 2$) and yields two mobility edges, one close to the Dirac point and another at a higher energy and dependent on the disorder strength. As a consequence, it opens the door for graphene based devices, which show an arbitrarily large {\it on/off} conductivity ratio.

The author acknowledges financial support from NSERC, FQRNT, RQMP and INTRIQ.


\begin{thebibliography}{99}

\bibitem{novo04} K. S. Novoselov, A. K. Geim, S. V. Morozov, D. Jiang, Y. Zhang, S. V. Dubonos, I. V. Grigorieva, and A. A. Firsov, Science \textbf{306}, 666 (2004).
\bibitem{geim07} A. K. Geim and K. S. Novoselov, Nat. Mater. \textbf{6}, 183 (2007).
\bibitem{CastroRev} A. H. Castro Neto, F. Guinea, N. M. R. Peres, K. S. Novoselov, A. K. Geim, Reviews of Modern Physics, \textbf{81}, 109 (2009). 
\bibitem{novo05} K. S. Novoselov, A. K. Geim, S. V. Morozov, D. Jiang, M. I. Katsnelson, I. V. Grigorieva, S. V. Dubonos, A. A. Firsov, Nature \textbf{438}, 197 (2005).
\bibitem{Mac06} K. Nomura, A.H. MacDonald, Phys. Rev. Lett, \textbf{96}, 256602 (2006); K. Nomura and A.H. MacDonald, Phys. Rev. Lett. \textbf{98}, 076602 (2007).
\bibitem{Castro06} J. Nilsson, A.H. Castro Neto, F. Guinea, and N.M.R. Peres, Phys. Rev. Lett., \textbf{97}, 266801 (2006).
\bibitem{Anderson58} P.W. Anderson, Phys. Rev. \textbf{109}, 1492 (1958).
\bibitem{gang4} E. Abrahams, P.W. Anderson, D.C. Licciardello, and T.V. Ramakrishnan, Phys. Rev. Lett. \textbf{42}, 673 (1979).
\bibitem{3Da} A. MacKinnon and B. Kramer, Phys. Rev. Lett. \textbf{47} 1546 (1981).
\bibitem{3Db} J.L. Pichard and G. Sarma, J. Phys. C: Solid State Phys. \textbf{14} L127 (1981).
\bibitem{3Dc} B. Kramer and A. MacKinnon, Rep. Prog. Phys. \textbf{56} 1469 (1993).
\bibitem{Schreiber92} M. Schreiber and M. Ottomeier, J. Phys.: Condens. Matter \textbf{4} 1959 (1992). 
\bibitem{Hilke03} M. Hilke, Phys. Rev. Lett., \textbf{91}, 226403 (2003).
\bibitem{Wallace47} P. K. Wallace, Phys. Rev. \textbf{71}, 622 (1947).
\bibitem{ishig07} M. Ishigami, J-H. Chen, W. G. Cullen, M. S. Fuhrer, and E. D. Williams, Nano Lett. \textbf{7}, 6 (2007).
\bibitem{chen08-2} J.-H. Chen, C. Jang, S. Xiao, M. Ishigami, and M. S. Fuhrer, Nature Nanotech. \textbf{3}, 206 (2008).
\bibitem{hwang07} E. H. Hwang, S. Adam, and S. Das Sarma, Phys. Rev. Lett. \textbf{98}, 186806 (2007).
\bibitem{graphane} D.C. Elias, R.R. Nair, T.M.G. Mohiuddin, S.V. Morozov, P. Blake, M.P. Halsall, A.C. Ferrari, D.W. Boukhvalov, M.I. Katsnelson, A. K. Geim, K. S. Novoselov, Science, \textbf{323}, 610 (2009).
\bibitem{Stormer08} K. I. Bolotin K. J. Sikes, Z. Jiang, M. Kilma, G. Fudenberg, J. Hone, P. Kim, and H. L. Stormer, Solid State Communications \textbf{146}, 351-355 (2008).
\bibitem{Yu09} V. Yu and M. Hilke, Appl. Phys. Lett. \textbf{95}, 151904 (2009).  
\bibitem{2D} I.Kh. Zharekeshev, M. Batsch and B. Kramer, Europhys. Lett. \textbf{34}, 587 (1996).
\bibitem{Hilke08} M. Hilke, Phys. Rev. B \textbf{78}, 012204 (2008).
\bibitem{UCF1} B.L. Altshuler, P.A. Lee, and R.A. Webb (ed) Mesoscopic Phenomena in Solids (Amsterdam: North-Holland) (1991).
\bibitem{UCF2} P.A. Lee and A.D. Stone, Phys. Rev. Lett. \textbf{55}, 1622 (1985).
\bibitem{Pendry94} J.B. Pendry, Adv. Phys., \textbf{43}, 461 (1994)
\bibitem{Mirlin06} P.M. Ostrovsky, I.V. Gornyi, and A.D. Mirlin, Phys. Rev. B \textbf{74}, 235443 (2006). 
\bibitem{Brouwer07} J. H. Bardarson, J. Tworzydlo, P.W. Brouwer, and C.W.J. Beenakker, Phy. Rev. Lett. \textbf{99}, 106801 (2007). 
\bibitem{Castro09} E.R. Mucciolo, A.H. Castro Neto, C.H. Lewenkopf, Phys. Rev. B \textbf{79} 075407 (2009). 
\bibitem{Roemer04} R.A. Römer, H. Schulz-Baldes, Europhys. Lett. \textbf{68}, 247 (2004). 
\bibitem{Nomura07} K. Nomura, M. Koshino and S. Ryu, Phys. Rev. Lett. \textbf{99}, 146806 (2007). 
\bibitem{Bostwick09} A. Bostwick, J.L. McChesney, K.V. Emtsev, T. Seyller, K. Horn, S.D. Kevan, and E. Rotenberg, Phys. Rev. Lett. \textbf{103}, 056404 (2009). 
\bibitem{Zhou08} S.Y. Zhou, D.A. Siegel, A.V. Fedorov, and A. Lanzara, Phys. Rev. Lett. \textbf{101}, 086402 (2008). 
\bibitem{Naumis07} Gerardo G. Naumis, Phys. Rev. B \textbf{76}, 153403 (2007). 
\bibitem{Amini09} M. Amini1, S.A. Jafari and F. Shahbazi1, Euro Phys. Lett., \textbf{87} 37002 (2009). 
\bibitem{Fehske09} G. Schubert, J. Schleede, and H. Fehske, Phys. Rev. B \textbf{79}, 235116 (2009). 

\end{thebibliography}
\end{document}